\documentclass[reqno,tbtags,intlimits,a4paper,oneside,12pt]{amsart}

\usepackage[foot]{amsaddr}

\usepackage[english]{babel}
\usepackage{amssymb,upref,calc,url}
\usepackage[mathscr]{eucal}
\usepackage{graphicx}
\usepackage{amsmath, amsfonts,amscd,amsthm,amssymb,verbatim}
\usepackage[all]{xy}
\usepackage{textcomp}
\usepackage[in]{fullpage}

\usepackage[title]{appendix}

\newtheorem{theorem}{Theorem}[section]
\newtheorem{corollaryt}[theorem]{Corollary}
\newtheorem{lemma}[theorem]{Lemma}
\theoremstyle{definition}
\newtheorem{definition}[theorem]{Definition}
\newtheorem{remark}[theorem]{Remark}

\usepackage{color}

\definecolor{shamrockgreen}{rgb}{0.0, 0.62, 0.38}

\usepackage{hyperref}
\hypersetup{
    colorlinks=true,
    linkcolor=shamrockgreen,
    filecolor=magenta,
    citecolor=blue,
    urlcolor=blue,
    pdftitle={The Mumford Dynamical System and the Gelfand--Dikii recursion},
    pdfpagemode=FullScreen,
    }

\title{The Mumford Dynamical System\\ and the Gelfand--Dikii recursion}

\author{Polina Baron}

\address{University of Chicago, USA}
\email{pbaron@uchicago.edu}

\begin{document}

\frenchspacing




\maketitle

\rightline{\em In memory of Israel Moiseevich Gelfand (1913--2009)}
\vspace{.5cm}

\begin{abstract}
In his paper \cite{Buch23}, Victor Buchstaber developed the differential-algebraic theory of the Mumford dynamical system.
The key object of this theory is the $(P,Q)$-recursion introduced in his paper.

In the present paper, we further develop the theory of $(P,Q)$-recursion and describe its connections to the Korteweg–-de Vries hierarchy, the
Lenard operator, and the Gelfand--Dikii recursion.\\
\\
------------------------------------------------------------------------------------------------------------------
\end{abstract}

\tableofcontents

\section*{Introduction}
The \emph{Korteweg–-de Vries equation (KdV equation)} on the function $u=u(x,\,t)$ is a nonlinear third-order partial
differential equation, classically written as
\begin{equation*}
4\frac{\partial u}{\partial t}=6uu'-u''',\quad\text{where }\,u'=\partial u = \frac  {\partial u}{\partial x}\,.
\end{equation*}

This equation describes a well-known and fundamental mathematical model of non-linear waves in dispersive media,
which is widely used to describe solitons, model turbulence in fluids, study the dispersion of light in optical fibers,
and to solve many other problems in classical and modern physics (see, for example, \cite{waves77}).
Exploration of solutions to the KdV equation and its connections to other fundamental equations of mathematical physics
led to the discovery of the \emph{Korteweg--de Vries hierarchy (KdV hierarchy)}: an infinite sequence of compatible partial
differential equations for the function $u=u(t_{0}=x,t_{1}=t;t_{2},t_{3},\dots)$ that has the form
\begin{align}
\label{KdVhierarchy1}
\partial_{k}U_1&=\partial U_{k+1},\\
\label{KdVhierarchy2}
\partial U_{k+1}&=\Lambda\partial U_{k}
\end{align}
for all $k\in\mathbb{N}$, where $\partial=\frac{\partial}{\partial t_0}=\frac{\partial}{\partial x}$, $\partial_{i}=\frac{\partial}{\partial t_i}$,
$i\geq 1$, the operator $\Lambda=\frac{1}{4}\partial^{2}-u-\frac{1}{2}u'\partial^{-1}$ is the pseudo-differential Lenard operator (see Definition \ref{lenarddef}),
$U_0$ is a constant, $U_1$ is a function in $u$, $\{U_i\}_{i\geq 2}$ are differential polynomials in $U_1$,
and equations \eqref{KdVhierarchy1}--\eqref{KdVhierarchy2} for $k=1$ yield the KdV equation.

The KdV hierarchy theory can be divided into two interrelated directions:

(1) The analytical direction, which targets construction and study of solutions to this hierarchy.

(2) The differential-algebraic direction, which aims to construct and study hierarchies whose solutions lead to solutions
of the KdV hierarchy and investigate the connections between the KdV hierarchy and other fundamental equations and hierarchies
of mathematical physics and integrable systems.

Some of the key results in the differential-algebraic direction were obtained by Gelfand and Dikii (see \cite{GD75} and \cite{GD79}).
One of these results is the construction of the \emph{the Gelfand--Dikii recursion} (see Definition \ref{GDdef}),
which provides a differential-algebraic solution to the KdV hierarchy in terms of the asymptotic resolvent of the Sturm--Liouville equation.
This resolvent is equivalent to the resolvent of the Schrödinger equation in quantum mechanics.

The \emph{Mumford dynamical system}, introduced in the book \cite[{Chapter 3, \S\,3, Theorem 3.1}]{Mumf}, is the dynamical system on the space $\mathbb{C}^{3g+1}$ with coordinates
$$
\mathbf{u}=(u_1,\,\ldots,\,u_g),\quad \mathbf{v}=(v_1,\,\ldots,\,v_g),\quad \mathbf{w}=(w_1,\,\ldots,\,w_{g+1})
$$
dependent on $\mathbf{t}=(t_1,\,\ldots,\,t_g)\in \mathbb{C}^{g}$ and
governed by the equations
\begin{align*}
\mathcal{D}_\eta u_\xi&= \frac{2}{\xi-\eta}(v_\xi u_\eta-u_\xi v_\eta), \\
\mathcal{D}_\eta v_\xi&=\frac{1}{\xi-\eta}(u_\xi w_\eta - w_\xi u_\eta) + u_\xi u_\eta, \\
\mathcal{D}_\eta w_\xi&=\frac{2}{\xi-\eta}(w_\xi v_\eta - v_\xi w_\eta) - 2v_\xi u_\eta.
\end{align*}
Here $\xi$ and $\eta$ are independent parameters, $\mathcal{D}_\eta = \sum_{i=1}^g \eta^{g-i}\partial_i$, and
\begin{equation*}
u_\xi = \xi^g+\sum_{i=1}^g u_i\xi^{g-i},\quad v_\xi = \sum_{i=1}^g v_i\xi^{g-i},\quad w_\xi = \xi^{g+1}+\sum_{i=1}^{g+1} w_i\xi^{g+1-i}.
\end{equation*}
This dynamical system possesses $2g+1$ integrals $h_1,\,\ldots,\,h_{2g+1}$ whose generating function is
\begin{equation*}
H_\xi = u_\xi w_\xi + v_\xi^2 = \xi^{2g+1} + \sum\limits_{n=1}^{2g+1}h_n \xi^{2g+1-n}.
\end{equation*}
See \cite[Section 1]{Buch23} for more details.

The paper \cite{Buch23} provides the general solution of the Mumford dynamical system (see \cite[Theorem 5.5]{Buch23}).
A key result is the construction of the $(P,Q)$-recursion (see \cite[Theorem 5.1]{Buch23}, also Definition \ref{pqdef} below).
\vspace{.25cm}

This paper further develops the theory of the $(P,Q)$-recursion. We present and prove some of its important properties
(see Theorems \ref{B=L} and \ref{sumofBtheorem}) and show its connections to the Gelfand--Dikii recursion through an invertible change of variables (see Theorems \ref{BGD} and \ref{BGD2}). The results of this paper are used in \cite{Buch23} to demonstrate
that the general solution of the Mumford dynamical system yields a general solution to the KdV hierarchy (see \cite[Theorem 7.5]{Buch23},
also Theorem \ref{KdVtheorem} below).

\vspace{.5cm}

\textbf{Acknowledgments:}
The author of this paper expresses deep gratitude to Victor Buchstaber for posing the problem and for meaningful discussions of the paper's results.

\section{The $(P,Q)$-recursion and its transformation into the $B(u;h)$-recursion}

\begin{definition}[Buchstaber \cite{Buch23}]\label{pqdef}
For any infinitely differentiable function $u=u(x)$ and an infinite vector of parameters $h=(h_1,\,h_2,\,\ldots)$, there exists an infinite sequence of differential polynomials
\begin{align*}
P_k&= P_k(u_1;h) = P_k(u_1,\,u_1',\,\dots,\,u_1^{(2k-2)};\;h),\\
Q_k&= Q_k(u_1;h) = Q_k(u_1,\,u_1',\,\dots,\,u_1^{(2k-2)};\;h)
\end{align*}
for all $k\in\mathbb{N}_0=\mathbb{N}\cup\{0\}$ that is defined by the parameterized recursion
\begin{alignat*}3
P_0&=1,&\quad P_1&= u_1, &\quad P_2&= \frac{1}{4}(u_1''+6u_1^2-4h_1u_1+2h_2),\\
Q_0&=1,&\quad Q_1&= h_1-u_1, &\quad Q_2&= \frac{1}{4}(-u_1''-2u_1^2+2h_2),
\end{alignat*}
and for all $k>2$,
\begin{align*}
P_k&=\frac{1}{4}P_{k-1}'' - \frac{1}{2}\sum\limits_{i+j=k}P_iQ_j - \frac{1}{8}\sum\limits_{i+j=k-1}P_i'P_j' - \frac{1}{2}(h_1-2u_1)P_{k-1} + \frac{1}{2}h_k,\\
Q_k&=-\frac{1}{4}P_{k-1}'' - \frac{1}{2}\sum\limits_{i+j=k}P_iQ_j - \frac{1}{8}\sum\limits_{i+j=k-1}P_i'P_j' + \frac{1}{2}(h_1-2u_1)P_{k-1} + \frac{1}{2}h_k
\end{align*}
Following \cite{Buch23}, we will call this parametric recursion \emph{the $(P,Q)$-recursion}.
\end{definition}

Note a slight difference of our definition from the one presented in \cite{Buch23}: for the sake of convenience,
we introduce additional polynomials $P_0$ and $Q_0$. This alteration does not change any of the results in both this paper and \cite{Buch23}.

\begin{remark}
The parameters $\{h_i\}_{i\in\mathbb{N}}$ of the $(P,Q)$-recursion are algebraically independent.
To obtain a hierarchy of functions over the given field of scalars, one assigns values from this field to these parameters.
\end{remark}

Let us perform the reversible change of variables
\begin{equation}\label{variablechange}
u=-2u_1+h_1;\quad B_i=-8P_i,\quad A_i=-8Q_i.
\end{equation}
Then $u_1=-\frac{1}{2}u+\frac{1}{2}h_1$,\; $P_i=-\frac{1}{8}B_i$, \;$Q_i=-\frac{1}{8}A_i$.
We obtain a different parameterised recursion that is equivalent to the $(P,Q)$-recursion.

\begin{definition}
Let $u=u(x)$ be an infinitely differentiable function, and let $h=(h_1,\,h_2,\,\ldots)$be an infinite vector of parameters.
There exists an infinite sequence of differential polynomials
\begin{align*}
B_k&= B_k(u;h) = B_k(u,\,u',\,\dots,\,u^{(2k-2)};\;h),\\
A_k&= A_k(u;h) = A_k(u,\,u',\,\dots,\,u^{(2k-2)};\;h)
\end{align*}
for all $k\in\mathbb{N}_0=\mathbb{N}\cup\{0\}$,
that is defined by parametrical recursion:
\begin{equation*}
\begin{alignedat}3
B_0&=-8,&\quad B_1&=4u-4h_1,&\quad B_2&=u''-3u^2+2h_1u+h_1^2-4h_2,\\
A_0&=-8,&\quad A_1&=-4u-4h_1,&\quad A_2&=-u''+u^2-2h_1u-h_1^2-4h_2,
\end{alignedat}
\end{equation*}
and for all $k>2$,
\begin{equation}\label{BA}
\begin{aligned}
B_{k+1}&=B_{k+1}(u;h)=\frac{1}{4}B_{k}''-\frac{u}{2}B_{k}
+\frac{1}{16}\sum_{i=1}^{k}B_iA_{k+1-i}+\frac{1}{64}\sum_{j=1}^{k-1}B_j'B_{k-j}'-4h_{k+1},\\
A_{k+1}&=A_{k+1}(u;h)=-\frac{1}{4}B_{k}''+\frac{u}{2}B_{k}
+\frac{1}{16}\sum_{i=1}^{k}B_iA_{k+1-i}+\frac{1}{64}\sum_{j=1}^{k-1}B_i'B_{k-j}'-4h_{k+1}.
\end{aligned}
\end{equation}
\end{definition}

Note that the sequence $\{B_k(u;h)\}_{k\in\mathbb{N}}$ can be defined without the use of the sequence $\{A_k(u;h)\}_{k\in\mathbb{N}}$.
For all $j\geq 1$,
\begin{equation}\label{AfromB}
A_{j}=B_{j}-(B_{j}-A_{j})=B_{j}-\frac{1}{2}B_{j-1}''+uB_{j-1}.
\end{equation}
Therefore,
\begin{multline}\label{BnoA}
B_{k+1}=\frac{1}{4}B_{k}''-\frac{u}{2}B_{k}-4h_{k+1}+\frac{1}{16}\sum_{i=1}^{k}B_i(B_{k-i+1}+uB_{k-i})-\\
-\frac{1}{64}\sum_{i=1}^{k-1}(B_i''B_{k-i}-B_i'B_{k-i}'+B_iB_{k-i}'').
\end{multline}

Evidently, such a representation complicates the formula significantly, so we won't use it.
However, because of that, we will call this parametric recursion \emph{the $B(u;h)$-recursion}.

\begin{remark}
The choice of the factor $-8$, somewhat unexpected at first glance, is motivated by the form of KdV hierarchy,
see \eqref{KdVhierarchy1}--\eqref{KdVhierarchy2}. From the definition of this hierarchy, $U_1=-4u$. Indeed,
$$
-4\partial_1 u = \partial_1 U_1 =
U_2' = \frac{1}{4}U_{1}'''- uU_{1}' - \frac{1}{2}uU_{1}' = -u''' + 6 uu'.
$$
Therefore, it is necessary to put $U_0=-8$ in order for the equations \eqref{KdVhierarchy1}--\eqref{KdVhierarchy2} to hold for $k=0$:
$$
-4 u' = U_1' =\frac{1}{4}U_{0}'''- uU_{0}''' - \frac{1}{2}u'U_{0} = - \bigg(\frac{1}{2}U_0\bigg) u'.
$$
Thus, the construction of the $B(u;h)$-recursion with the initial condition $B_0=-8$ leads to a more convenient form of subsequent formulas (see, for example, Theorem \ref{KdVtheorem}).
\end{remark}

\section{Basic properties of the $B(u;h)$-recursion}

\subsection{Differential form of the $B(u,h)$-recursion and the Lenard operator}
\begin{definition}\label{lenarddef}
The \emph{Lenard operator} is a pseudo-differential operator
\begin{equation*}
\Lambda=\Lambda_{u(x)}
=\frac{1}{4}\partial^{2}-u-\frac{1}{2}u'\partial^{-1}
=\frac{1}{4}\frac{\partial^2}{\partial x^2}-u(x)-\frac{1}{2}u'(x)\bigg(\frac{\partial}{\partial x}\bigg)^{-1}.
\end{equation*}
\end{definition}

Note that the Lenard operator $\Lambda$ is not defined for constants and polynomials in $u$.
However, one can define the differential operator $\Lambda\partial$ on functions $f\in\mathbb{C}[u,\,u',\,u'',\,\dots]$ by the formula
\begin{equation*}
\Lambda\partial (f)=\frac{1}{4}f'''-uf'-\frac{1}{2}u'f.
\end{equation*}

\begin{theorem}\label{B=L}

For all $k\geq 0$, the $B(u;h)$-recursion satisfies the equation
\begin{equation}\label{lenardeq}
\partial B_{k+1} = \Lambda\partial B_{k}, \quad\text{or}\quad  B_{k+1}' = \frac{1}{4}B_k'''-uB_k'-\frac{1}{2}u'B_k.
\end{equation}
\end{theorem}

A reverse change of variables given by the equation \eqref{variablechange} leads to
\begin{corollaryt}
For all $k\geq 0$, the $(P,Q)$-recursion satisfies the equation
\begin{equation}\label{lenardeq}
P_{k+1}' = \frac{1}{4}P_k'''+(2u_1-h_1)P_k'-2u_1'P_k.
\end{equation}
\end{corollaryt}

To prove Theorem \ref{B=L}, we need the following technical result.

\begin{lemma}\label{sumlemma}
For all $k\geq 0$,
\begin{equation}\label{sumeq}
B_{k+1}'+A_{k+1}'=-uB_k'.
\end{equation}
\end{lemma}
\begin{proof}
We will prove this lemma via induction on $k$. For all $k\in\{0,1\}$, we have
$$
(B_{1}+A_{1})'=4u' -4u'=0=-uB_0',\quad (B_{2}+A_{2})'=(-2u^2)'=-4uu'=-uB_1'.
$$

Fix $k>2$. Suppose that Equation \eqref{sumeq} holds for all $0\leq i\leq k$. Let us prove that it should also hold for $i=k+1$.

By definition,
\begin{align*}
8(B_{k+1}+A_{k+1})'&=\sum_{i=1}^{k}(B_iA_{k-i+1})'+\frac{1}{4}\sum_{j=1}^{k-1}(B_i'B_{k-j}')'=\\
&=\sum_{i=1}^{k}(B_i'A_{k-i+1}+B_iA_{k-i+1}')+\frac{1}{2}\sum_{j=1}^{k-1}B_i'B_{k-j}''.
\end{align*}
Using the Equation \eqref{AfromB}, we can change the variables $A_{k+1-i}=B_{k-i+1}-\frac{1}{2}B_{k-i}''+uB_{k-i}$ for all $1\leq i\leq k-1$.
By the assumption of the induction, we can change the variables $A'_{k+1-i}=-uB_{k-i}'-B_{k-i+1}'$ for all $1\leq i\leq k-1$. Then
\begin{align*}
8(B_{k+1}+A_{k+1})'&=(B_kA_1)'+\sum_{i=1}^{n-1}\bigg(B_i'\bigg[B_{k-i+1}-\frac{1}{2}B_{k-i}''+uB_{k-i}\bigg]+\\[-5pt]
&\kern100pt+B_i[-uB_{k-i}'-B_{k-i+1}']\bigg)+\frac{1}{2}\sum_{i=1}^{k-1}B_i'B_{k-i}''=\\[-5pt]
&=(B_kA_1)'+\sum_{i=1}^{k-1}\bigg(\frac{1}{2}B_i'B_{k-i}''-B_i'B_{k+1-i}+B_iB_{k+1-i}'-\\[-5pt]
&\kern110pt-uB_i'B_{k-i}+uB_iB_{k-i}'\bigg)+\frac{1}{2}\sum_{i=1}^{k-1}B_i'B_{k-i}''=\\
&=(B_k'A_1+B_kA_1')+(B_k'B_1-B_1'B_k)=\\
&=B_k'(B_1-A_1)-B_k(A_1'+B_1')=-8uB_k'.
\end{align*}
\end{proof}

{\it Proof of Theorem \ref{B=L}.}
Differentiating Equation \eqref{BA} for $B_{k+1}$, obtain
$$
B_{k+1}'=\frac{1}{16}\sum_{i=1}^{k}(B_iA_{k+1-i})'+\frac{1}{64}\sum_{i=1}^{k-1}(B_{i}'B_{k-i}')'
+\frac{1}{4}B_{k}'''-\frac{u'}{2}B_{k}-\frac{u}{2}B_{k}'.
$$
Subtracting Equation \eqref{lenardeq}, get
$$
0=\frac{1}{16}\sum_{i=1}^{k}(B_iA_{k+1-i})'+\frac{1}{64}\sum_{i=1}^{k-1}(B_{i}'B_{k-i}')'+\frac{u}{2}B_{k}'.
$$
Rearranging the terms in this equation and multiplying it by $2$ gives
$$
-uB_k'=\frac{1}{16}\sum_{i=1}^{k}(B_iA_{k+1-i})'+\frac{1}{64}\sum_{i=1}^{k-1}(B_{i}'B_{k-i}')'=(B_{k+1}+A_{k+1})',
$$
which is exactly the statement of Lemma \ref{sumlemma}.
\hfill$\Box$

\subsection{Solving the $B(u,h)$-recursion in terms of the $\mathbf{B}(u)$-recursion} \text{}\\
In the $B(u;h)$\-recursion, the function $u$ and the vector of parameters $h=(h_1,\,h_2,\,\ldots)$ are independent.
Therefore, we can consider the $B(u;h)$-recursion under various constraints, including $u\equiv0$ and $h=(h_1,\,h_2,\,\ldots)=(0,\,0,\,\ldots)=0$.

Consider the sequence $(B_i,A_i)$, $i\in\mathbb{N}_0,$ with constraint $u\equiv0$. We get
\begin{gather*}
B_0(0;h)=A_0(0;h)=-8,\quad B_1(0;h)=A_1(0;h)=-4h_1,\\
B_2(0;h)=A_2(0;h)=-4h_2+h_1^2,
\end{gather*}
and for any $k\geq 3$,
$$
B_k(0;h)=-4h_k+\frac{1}{16}\sum_{i=1}^{k-1}B_i(0;h)A_{k-i}(0;h)=A_k(0;h).
$$
Since $B_k(0;h)=A_k(0;h)$ for all $k\in\mathbb{N}_0$, 
$$
B_k(0;h)=-4h_k+\frac{1}{16}\sum_{i=1}^{k-1}B_i(0;h)B_{k-i}(0;h).
$$
Since $B_0(u;h)=-8$ for any $u$ and $h$, 
$$
B_0(0;h)B_k(0;h)+B_k(0;h)B_0(0;h)=-16B_k(0;h).
$$
Then,
$$
h_k=\frac{1}{4}\sum_{i=0}^{k}B_i(0;h)B_{k-i}(0;h).
$$

Setting $\beta_i=\beta_i(h)=-\frac{1}{8}B_i(0;h)$ for all $i\geq 0$, we obtain a uniquely defined infinite recurrent sequence of parameters
\begin{equation}\label{beta-rec}
\beta_0=1,\qquad
\beta_1=\frac{1}{2}h_1,\quad\quad
\beta_k=\frac{1}{2}h_k-\sum_{i=1}^{k-1}\beta_i\beta_{k-i}\quad\text{for }\,k>1.
\end{equation}

\begin{definition}
\emph{The $\mathbf{B}(u)$-recursion} is the $B(u;h)$-recursion with $h_i=0$ for all $i\in\mathbb{N}$. Explicitly, for all $k\in\mathbb{N}_0$,
$$
\mathbf{B}_k=\mathbf{B}_k(u)=B_k(u;0).
$$
\end{definition}

Note that
\begin{gather*}
B_0=-8=\beta_0\mathbf{B}_0,\qquad B_1=4u-4h_1=\beta_0\mathbf{B}_1+\beta_1\mathbf{B}_0, \\
B_2=u''-3u^2+2h_1u+h_1^2-4h_2=\beta_0\mathbf{B}_2+\beta_1\mathbf{B}_1+\beta_2\mathbf{B}_0.
\end{gather*}

\begin{theorem}\label{sumofBtheorem}
For all $k>0$,
\begin{equation}\label{sumofB}
B_k=\sum_{j=0}^{k}\beta_{k-j}\mathbf{B}_j,
\end{equation}
where the coefficients $\{\beta_i\}_{i\in\mathbb{N}_0}$ are defined by the recursion \eqref{beta-rec}.
\end{theorem}

\begin{proof}
We use induction on $k$. As we have just shown, the statement of the theorem holds for $k\in\{0,1,2\}$.
Fix $k\geq 2$ and assume that Equation \eqref{sumofB} holds for any $0\leq i\leq k$.
Let us use this assumption to change variables in Equation \eqref{lenardeq}. We obtain
\begin{align*}
B_{k+1}'&=\frac{1}{4}B_k'''-uB_k'-\frac{1}{2}u'B_k=\\
&=\frac{1}{4}\bigg(\sum_{j=0}^{k}\beta_{k-j}\mathbf{B}_j\bigg)'''
-u\bigg(\sum_{j=0}^{k}\beta_{k-j}\mathbf{B}_j\bigg)'-\frac{1}{2}u'\bigg(\sum_{j=0}^{k}\beta_{k-j}\mathbf{B}_j\bigg)=\\
&=\sum_{j=0}^{k}\beta_{k-j}\bigg(\frac{1}{4}\mathbf{B}_j'''-u\mathbf{B}_j'-\frac{1}{2}u'\mathbf{B}_j\bigg)
=\sum_{j=0}^{k}\beta_{k-j}\mathbf{B}_{j+1}'=\sum_{j=1}^{k+1}\beta_{k+1-j}\mathbf{B}_{j}'.
\end{align*}
Therefore, there exists a constant $C$ such that
$$
B_{k+1}=\sum_{j=1}^{k+1}\beta_{k+1-j}\mathbf{B}_{j}+C.
$$
Evidently, $C=B_{k+1}(0;h)=-8\beta_{k+1}$. Since $B_0(u;h)=-8$ for any $u$ and $h$, we can express $C$ in the form
$$
C=\beta_{k+1}\mathbf{B}_{0}.
$$
Hence, the statement of the theorem holds for $k+1$.
\end{proof}

\begin{corollaryt}
For all $k>0$, 
\begin{equation*}
P_k=\sum_{j=0}^{k}\beta_{k-j}\mathbf{P}_j,
\end{equation*}
where $\mathbf{P}_j=P_j(u_1;0)\;\,\forall\, j\in\mathbb{N}_0$, and the coefficients $\{\beta_i\}_{i\in\mathbb{N}_0}$ are defined by Equation \eqref{beta-rec}.
\end{corollaryt}

\section{Solving the Gelfand-Dikii recursion in terms of the $B(u,h)$-recursion}
\begin{definition}\label{GDdef}
Consider a formal series
$$
R=\frac{1}{z}\sum_{k=0}^{\infty}R_{k}z^{-2k},\qquad R_{k}\in\mathbb{C}[u,u',u'',\dots],\;k\in\mathbb{N}_0,
$$
such that
\begin{equation}\label{GDeq}
-2RR''+(R')^2+4(u+z)R^2 = c(z)=1+\sum_{k=1}^{\infty}c_{k}z^{-2k},
\end{equation}
where $c_{k}$ are arbitrary constants. The recursion of the sequence of polynomials $\{R^c_{k}\}_{k\in\mathbb{N}_0}$
that solves Equation \eqref{GDeq} is called \emph{the Gelfand--Dikii recursion (GD-recursion)}.
The sequence of polynomials $\{R_{k}\}_{k\in\mathbb{N}_0}$ that solves Equation \eqref{GDeq} with constraint $c_{k}=0$ for all $k\in\mathbb{N}$
is called \emph{the standard solution of the Gelfand--Dikii recursion}.
\end{definition}

Evidently, for any series $c(z)$ and its corresponding solution $R^c$, we have
\begin{equation*}
R^c=\alpha(z)R,\qquad \text{where }\;\alpha(z)=1+\sum_{k=1}^{\infty}\alpha_{k}z^{-2k},\;\;\alpha_{k}\,\text{are constant,\;\; and}\; \alpha(z)^2=c(z).
\end{equation*}

Gelfand and Dikii showed (see \cite[Chapter 2, Equation (9)]{GD75}; also see \cite[\S 1, Equation (1.3)]{GD79})
that the standard solution of the GD-recursion satisfies the equation
\begin{equation}\label{GDL}
\partial R_{k+1} = \Lambda\partial R_{k},\;\;\text{ or }\;R_{k+1}' = \frac{1}{4}R_{k}''' - uR_{k}' - \frac{1}{2}u'R_{k}.
\end{equation}
Moreover, they provided (see \cite[Chapter 2, Equation (8)]{GD75}; see also \cite[\S 1, Equation (1.4)]{GD79})
an explicit recurrent formula for the standard solution. In our notation, with a minor correction to Formula (8) from \cite[Chapter 2]{GD75}, this formula is
\begin{equation}\label{GDformula}
\begin{aligned}
R_0=\frac{1}{2}, &\qquad R_1=-\frac{1}{4}u,\\
R_{k+1}=\frac{1}{2}\sum_{i=0}^{k-1}R_{i}R_{k-i}''-\frac{1}{4}\sum_{i=1}^{k-1}R_{i}'R_{k-i}'&-u\sum_{i=0}^{k}R_{i}R_{k-i}-\sum_{i=1}^{k}R_{i}R_{k-i+1}
\; \text{for all }\,k\in\mathbb{N}.
\end{aligned}
\end{equation}

\begin{remark}\label{GDremark}
Equation \eqref{GDformula} can be rewritten in the form
$$
R_{k+1} = \frac{1}{4}\sum_{i=1}^{k-1}\big(R_{i}R_{k-i}''-R_{i}'R_{k-i}'+R_{i}''R_{k-i}\big)-\sum_{i=1}^{k}R_{i}\big(R_{k-i+1}+uR_{k-i}\big)+\frac{1}{4}R_{k}''-\frac{1}{2}uR_{k}.
$$
The substitution $B_i=-16 R_i, \; i\in\mathbb{N}_0,$ gives Equation \eqref{BnoA} for $h=0$.
\end{remark}

A more detailed definition of the Gelfand--Dikii recursion and a derivation of Equation \eqref{GDL} are given in \cite[Section 6]{Buch23}.
\vspace{.5cm}

Note that $R_0(u)=\frac{1}{2} = -\frac{1}{16}\mathbf{B}_0(u)$. From Theorem \ref{B=L} and Lemma 7.1 in \cite{Buch23}, or from Remark \ref{GDremark}, 

\begin{theorem}\label{BGD}
  The solution of the $-\frac{1}{16}\mathbf{B}(u)$-recursion coincides with the standard solution of the Gelfand--Dikii recursion.
\end{theorem}

Furthermore, from Theorem \ref{BGD} and Theorem \ref{sumofBtheorem},

\begin{theorem}\label{BGD2}
  The solution of the $-\frac{1}{16}B(u;h)$-recursion with the parameters $\{h_{i}\}_{i\in\mathbb{N}}$ coincides with the solution of the general Gelfand--Dikii recursion with parameters $\{c_{i}\}_{i\in\mathbb{N}}$ if and only if
  \begin{equation}\label{h2c}
c_{i}=h_i \quad\text{for all }\,i\in\mathbb{N}.
\end{equation}
\end{theorem}

\begin{corollaryt}
The solution of the $(P,Q)(u_1,\,h)$-recursion with the parameters $\{h_{i}\}_{i\in\mathbb{N}}$ coincides with the solution of the general Gelfand--Dikii\; $-2R(-2u_1+h_1)$-recursion with the parameters $\{c_{i}\}_{i\in\mathbb{N}}$ if and only if Equation \eqref{h2c} holds.
\end{corollaryt}

\begin{remark}
  The condition of Equation \eqref{h2c} is equivalent to the condition
\begin{equation*}
\alpha_{i}=\beta_i=-\frac{1}{16}B_i(0;h)=\frac{1}{2}P_i(h_1;h)\quad \text{for all }\,i\in\mathbb{N}.
\end{equation*}
\end{remark}

From Theorem \ref{BGD2}, properties of the recursion, and Lemma 7.1 in \cite{Buch23},
\begin{theorem} \label{KdVtheorem}
The ${B}(u)$-recursion is a special solution to the Korteweg--de Vries hierarchy.
Moreover, the ${B}(u;h)$-recursion solves the Korteweg--de Vries hierarchy for any vector of parameters $h$.
\end{theorem}

\begin{corollaryt}[see also {\cite[Theorem 7.5]{Buch23}}]\label{KdVcor}
The solution of the $-8(P,Q)(-\frac{1}{2}u_1,0)$-recursion is a special solution of the Korteweg--de Vries hierarchy.

Moreover, the solution of the $-8(P,Q)(-\frac{1}{2}u_1+\frac{1}{2}h_1,\,h)$-recursion is a solution of the general Korteweg--de Vries hierarchy for any vector of parameters $h$.
\end{corollaryt}

\vspace{1cm}


\end{document}